\documentclass[a4paper,12pt]{article}
\usepackage{amsfonts}
\usepackage{graphicx}
\usepackage{amsmath}
\usepackage{color}
\usepackage{cite}

\newcommand{\be}{\begin{equation}}
\newcommand{\ee}{\end{equation}}
\newcommand{\bpartial}[0]{\bar{\partial}}
\newcommand{\bgamma}[0]{\bar{\gamma}}
\newcommand{\bbeta}[0]{\bar{\beta}}

\begin{document}


\begin{center}
\noindent{{\LARGE{$T\bar{T}$ type deformation in the presence of a boundary}}}
\smallskip
\smallskip

\smallskip
\smallskip

\smallskip
\smallskip

\smallskip
\smallskip

\noindent{{Juan Pablo Babaro$^{a}$, Valentino F.\ Foit$^{b}$, Gaston Giribet$^{b}$, Matias Leoni$^{a}$}}

\smallskip
\smallskip

\smallskip
\smallskip

\smallskip
\smallskip

\smallskip
\smallskip

\smallskip
\smallskip

$^a${Departamento de F\'{\i}sica, Universidad de Buenos Aires \& IFIBA - CONICET}\\ 
{\it Ciudad Universitaria, pabell\'on 1 (1428) Buenos Aires, Argentina}

\smallskip

$^b${Center for Cosmology and Particle Physics, Physics Department, New York University}\\
{\it 726 Broadway, New York, NY 10003, USA}


\end{center}

\smallskip
\smallskip

\smallskip
\smallskip

\smallskip
\smallskip

\smallskip

\begin{center}
{\bf Abstract}
\end{center}
We continue the study of a recently proposed solvable irrelevant deformation of an AdS$_3$/CFT$_2$ correspondence that leads in the UV to a theory with Hagedorn spectrum. This can be thought of as a single trace analog of the $T\bar{T}$-deformation of the dual CFT$_2$. Here we focus on the deformed worldsheet theory in presence of a conformal boundary. First, we compute the expectation value of a bulk primary operator on the disc geometry. We give a closed expression for such observable, from which we obtain the anomalous conformal dimension induced by the deformation. We compare the result with that coming from the computation of the 2-point correlation function on the sphere, finding exact agreement. We perform the computation using different techniques and making a comparative analysis of different regularization schemes to solve the logarithmically divergent integrals. This enables us to perform further consistency checks of our result by computing other observables of the deformed theory: We compute both the bulk-boundary 2-point and the boundary-boundary 2-point functions and are able to reproduce the anomalous dimensions of both boundary and bulk operators.



\newpage

\section{Introduction}

It was recently shown in \cite{SmirnovZamolodchikov, Cavaglia} that deforming a general two-dimensional conformal field theory (CFT$_2$) by adding to its action the irrelevant operator $T\bar{T}$, where $T$ refers to the holomorphic component of the stress-tensor \cite{Zamolodchikov}, retains integrability properties and defines a solvable QFT. This is a very important result in the study of the renormalization group flow and this is the reason why it attracted much attention recently \cite{TT1, TT2, TT3, TT4, TT5, TT6, TT7, TT8, TT9, TT10, TT11, TT12, TT13,Baggio:2018gct,Dei:2018mfl}. In particular, this modification was studied in the context of holography in \cite{Verlinde}, where it was proposed that the UV deformation is geometrically realized by a cutoff that removes the asymptotic region of AdS$_3$ space and replaces it by wall at finite distance from the boundary, where a QFT with Dirichlet boundary conditions is defined. 

As pointed out in \cite{GIK}, in the context of holography, a large class of solvable deformed CFTs can be obtained by studying string theory on AdS$_3$ with NS-NS fluxes. There it was shown that a single-trace analog of the $T\bar{T}$ deformation of the boundary CFT$_2$ gives rise in the bulk to string theory in a background that interpolates between AdS$_3$ in the IR and a linear dilaton background of Little String Theory in the UV. This represents quite an interesting setup, which raises the hope to work out the details of a non-AdS holography scenario.

The irrelevant deformation studied in \cite{GIK} shares some qualitative features with the original $T\bar{T}$-deformation of \cite{SmirnovZamolodchikov, Cavaglia}, in particular, the property of being solvable and universal \cite{GIK2}. The model of \cite{GIK}, however, follows from a rather different approach. It is based on the worldsheet formulation of the bulk theory, and the marginal deformations, when interpreted from the dual point of view, give rise to an irrelevant deformation in the boundary. In fact, the model can be regarded as a single trace version of $T\bar{T}$. This yields a solvable deformation of AdS$_3$/CFT$_2$ duality, which leads to a theory with a Hagedorn entropy in the UV. The spectrum of the theory can be explicitly obtained and compared with the spectrum predicted in \cite{SmirnovZamolodchikov, Cavaglia}; this was done in \cite{GIK2}. Correlation functions for the model of \cite{GIK} on the sphere topology were also computed \cite{GIK3, Gaston} which led to interesting observations about the theory, especially in relation to its non-locality. The analytic properties of the spectral density, the asymptotic convergence of the perturbation theory, and the anomalous dimensions induced by the deformation were analyzed. Other features, such as the structure of spatial entanglement and its comparison with the standard $T\bar{T}$ deformation, were also studied recently \cite{GIK4}.

In this paper, we will continue the study of this irrelevant deformation of AdS$_3$/CFT$_2$ by extending the results of \cite{Gaston} to the case in which the worldsheet theory has boundaries. More precisely, we will consider the marginal deformation of the worldsheet theory on AdS$_3$, as proposed in \cite{GIK}, formulated on the disc geometry with conformal symmetry preserving boundary conditions. In the undeformed theory, this describes AdS$_2$ D-branes in terms of correlation functions on the disc. For the deformed theory these observables have not yet been computed, and this is the computation we undertake in this paper. In section 2, we will review the bulk theory in presence of the deformation, as proposed in \cite{GIK}. In section 3, we will discuss the contributions to the action coming from the boundary, which amounts to discussing the appropriate boundary conditions. In section 4, we present the correlation functions we want to compute and our strategy for obtaining the anomalous dimensions induced by the deformation of the theory. In section 5, we compute the 1-point function of a bulk primary operator in the disc geometry. We obtain the expression for the anomalous dimension, which we compare with the result coming from the sphere 2-point function. In section 6, we do a similar computation but involving two operators inserted in the boundary of the disc. Using path integral techniques, we compute the boundary-boundary 2-point function in the deformed theory in terms of the analogous observable for the Wess-Zumino-Witten (WZW) model. In section 7, we compute the much more involved bulk-boundary 2-point function, which gives a non-trivial consistency check of the results obtained in the previous sections. As a further consistency check, in section 8 we reproduce the results for the anomalous dimensions using perturbation theory. We include three appendices with the details of the calculations.

\section{Bulk theory and IR deformation}

The bulk theory is defined by an action of the form\footnote{When comparing with \cite{Gaston}, consider the changes in conventions: $\phi \to -\phi/\sqrt{2}$, $M_0 \to 2 M_0/b^{2}$, $J^-\to J^+$.} $S=S_{\text{WZW}}+S_{\text{D}}+S_{b}$ consisting of a level $k=2+b^{-2}$ $SL(2,\mathbb{R})$ WZW theory action
\begin{equation}
S_{\text{WZW}}=\frac{1}{2\pi}\int\limits_\Gamma d^2z\, g^{1/2}
\left(\partial\phi\bar\partial\phi+\beta\bar\partial\gamma+\bar\beta\partial\bar\gamma
+\frac{b}{4}R\phi-b^2M_0\beta\bar\beta e^{2b\phi}
\right) , \label{TheS}
\end{equation}
deformed by a marginal operator
\begin{equation}
S_{{D}}=-\frac{\lambda_0}{\pi}\int\limits_{\Gamma}d^2z\,g^{1/2}\beta\bar\beta, \label{SD}
\end{equation} 
and a boundary action $S_b$, which we will discuss in the next section. $\Gamma$ is the Riemann surface corresponding to the disc geometry, which can be mapped to the complex upper half plane. More precisely, $\Gamma $ will be taken to be the upper half plane, i.e.\ $y\geq 0$ with $z=x+iy$, while the boundary will be given by the real line $z=x$. 

Bulk action $S=S_{\text{WZW}}+S_{{D}}$ has been studied in detail in \cite{GIK, GIK2, Gaston, GIK3, GIK4}, and it appeared in the literature before in different contexts; see for instance \cite{Israel}. In the case $\lambda_0=0$ it corresponds to the $SL(2,\mathbb{R})$ WZW model, which describes the string $\sigma$-model on AdS$_3$; see \cite{GKS} and references therein and thereof. In presence of the deformation (i.e.\ $\lambda_0\neq 0$) it describes a string geometry that interpolates between AdS$_3$ and a linear dilaton background. Indeed, $S_D$ represents a worldsheet marginal deformation, which is build up by two of the $SL(2,\mathbb{R})$ Kac-Moody currents of the WZW model, i.e.\ $J^- \bar{J}^-=\beta\bar{\beta }$. This makes the deformation to be universal, in the sense that it will be present in all AdS$_3\times M$ string theory backgrounds with affine symmetry. The deformation (\ref{SD}) does break $SL(2,\mathbb{R})$ symmetry but is exactly marginal in the sense that it preserves conformal invariance. The deformation is still solvable in the sense that the spectrum and correlation functions can be exactly derived for finite $\lambda_0$ \cite{GIK2, GIK3, Gaston}. 

We already mentioned that (\ref{SD}) can be though of as inducing a single-trace version of the $T\bar{T}$-deformation in the dual theory. To see this, one can consider the stress-tensor of the boundary CFT$_2$ dual to the AdS$_3$ string theory. The general form of such a tensor was obtained in \cite{KutasovSeiberg}, where it was shown to be given as the worldsheet integration of certain local fields, namely
\begin{equation}
T=\int d^2z (\partial_{\text{x}} J\partial_{\text{x}} +2\partial_{\text{x}}^2J)\Phi_1 \bar{J},
\end{equation} 
with $\partial_{\text{x}}$ being the derivative with respect to an auxiliary complex variable ${\text{x}}$ that organizes the $SL(2,\mathbb{R})$ representations. $J$ is composed by the three Kac-Moody local currents $J^{3,\pm }$ written as a polynomial in ${\text{x}}$, and $\Phi_1 $ is a bulk primary field dual to a boundary operator of conformal dimension $1$; see \cite{KutasovSeiberg, GIK} for details. An analogous expression holds for the anti-holomorphic counterpart $\bar{T}$ by replacing ${\partial }_{{\text{x}}}\leftrightarrow {\partial }_{\bar{\text{x}}}$ and $ J\leftrightarrow \bar{J}$. This leads to define the boundary $T\bar{T}$ operator as follows
\begin{equation}
S_{T\bar{T}}=-\int d^2{\text{x}} \int d^2z \ (\partial_{\text{x}} J\partial_{\text{x}} +2\partial_{\text{x}}^2J)\Phi_1 \bar{J}\ \cdot \ c.c.\label{ElTT},
\end{equation}
since from the boundary CFT$_2$ perspective, the variable ${\text{x}}\in \mathbb{C}$ represents the coordinates where the dual operators are inserted. $c.c.$ stands for the complex conjugate part, defined by an independent integration over the worldsheet variable $z'\in \mathbb{C}$ of the analogous expression obtained by replacing ${\text{x}}\leftrightarrow \bar{{\text{x}}}$, $J(z)\leftrightarrow \bar{J}(\bar{z}')$, and $\bar{J}(\bar{z})\leftrightarrow J(z')$. 

Operator (\ref{SD}), in contrast, is given by a similar but different formula, namely
\begin{equation}
S_D=-\int d^2{\text{x}} \int d^2z \ (\partial_{\text{x}} J\partial_{\text{x}} +2\partial_{\text{x}}^2J) (\partial_{\bar{{\text{x}}}} \bar{J}\partial_{\bar{{\text{x}}}} +2\partial_{\bar{{\text{x}}}}^2\bar{J})\Phi_1 = -\frac{\lambda_0 }{\pi }\int d^2z \ J^-\bar{J}^-,\label{sT}
\end{equation} 
where $\lambda_0$ is defined by integrating $\Phi_1$ over the worldhseet, and boundary terms have been dismissed; see \cite{GIK, GIK2} for details. The single integration over the worldsheet variable in (\ref{sT}) explains in what sense this operator can be regarded as a single trace version of (\ref{ElTT}).

\section{Boundary theory and boundary conditions}

Let us now discuss the boundary action $S_b$ which is given by
\begin{equation}
S_{{b}}=\frac{1}{4\pi}\int\limits_{\partial\Gamma}dx\ g^{1/4}\left(2b K\phi+{i}
\beta(\gamma+\bar\gamma)-{i\zeta}\beta e^{b\phi}-{i\lambda_b}\beta\right), \label{Slab}
\end{equation}
where $\zeta $ is an arbitrary constant. $\partial \Gamma$ refers to the boundary of $\Gamma$, i.e.\ the real line in the case of the upper half plane representation of the disc. The factor $g^{1/4}$ in the boundary integration measure stands for the Jacobian written in terms of the induced metric. Hereafter we will omit the factor and work in the conformal frame. We will mainly follow the conventions of \cite{FR, Babaro}. The boundary action $S_b$ contains the boundary terms proposed in \cite{FR} together with an additional term 
\begin{equation}
-\frac{i\lambda_b}{4\pi}\int_{\partial\Gamma} dx\beta \ .
\end{equation}

After integrating the $\beta\gamma$ fields in (\ref{Slab}) by parts, we obtain the action
\begin{eqnarray}
S &=& \frac{1}{2\pi}\int\limits_\Gamma d^2z\, 
\left(\partial\phi\bar\partial\phi-\gamma\bar\partial\beta-\bar\gamma\partial\bar\beta
+\frac{b}{4}R\phi-b^2M_0\beta\bar\beta e^{2b\phi}-2\lambda_0\beta\bar\beta
\right)+\nonumber\\
&& \frac{1}{4\pi}\int\limits_{\partial\Gamma}dx 
\left(2b K\phi-{i\zeta}\beta e^{b\phi}-{i\lambda_b}\beta \right).
\end{eqnarray}

Considering the boundary terms in its variation, using the constraint $\delta(\beta+\bar\beta)|_{z=\bar z}=0$, we have
\begin{equation}
\delta S_{{b}}=\frac{i}{4\pi}\int\limits_{\partial\Gamma}dx
\Big(\delta\phi\left((\bar\partial-\partial)\phi-\zeta b\beta e^{b\phi}\right)+\delta\beta
\left(\gamma+\bar\gamma-\zeta e^{b\phi}-\lambda_b\right)\Big)
\end{equation}
from which we obtain the gluing conditions
\begin{eqnarray}
 \beta+\bar\beta|_{z=\bar z}=0 \ , \ \  \ (\bar\partial-\partial)\phi|_{z=\bar z}=\zeta b\beta e^{b\phi}\ , \ \ \ \gamma+\bar\gamma|_{z=\bar z}=\zeta e^{b\phi}+\lambda_b,  \label{gluing1}
\end{eqnarray}
valid at the boundary, where $z=\bar z$, as the subscript indicates. As we will discuss below, these gluing conditions are consistent with
\begin{eqnarray}
J^{-}+\bar J^{-}|_{z=\bar z}=0\ , \ \ \  T(z)-\overline T(\bar z)|_{z=\bar z}=0 \ . \label{gluing2}
\end{eqnarray}
The one on the left is the boundary condition of the only Kac-Moody current that is still conserved, i.e.\ $J^{-}=\beta$. The one on the right is the boundary condition of the worldsheet stress-tensor $T(z)=-\beta\partial\gamma-(\partial\phi)^2+b\partial^2\phi$. While the former follows immediately from the first condition in (\ref{gluing1}), the latter is more involved and requires to be proven. It will be enough to prove this at classical level\footnote{A more definite argument valid at quantum level would demand verifying the conditions (\ref{gluing2}) for fields inside correlators; see \cite{FR}.}. Consider the rescaled fields $2b\phi\to\varphi$, $b\beta\to\beta_{cl}$, and $b\gamma\to\gamma_{cl}$, the rescaled constant $b^2M_0=\tilde\lambda$, and the rescaled boundary parameters $b\zeta\to\tilde\zeta$ and $b\lambda_b\to\tilde\lambda_b$. Then, we define the classical limit of the stress-tensor components as 
\begin{align}
\begin{split}
T_{cl}(z)&=\lim\limits_{b^2\to 0}b^2 T(z)=-\beta_{cl}\partial\gamma_{cl}-\frac{1}{4}\left(\partial\varphi\right)^2,\\
\overline{T}_{cl}(\bar z)&=\lim\limits_{b^2\to 0}b^2 \overline{T}(\bar z)=-\bar\beta_{cl}\partial\bar\gamma_{cl}-\frac{1}{4}\left(\bar\partial\varphi\right)^2
\end{split}
\end{align}
which, with the use of the classical equations of motion,
\begin{align}\label{eom}
& \partial\bar\partial\varphi=-2\tilde\lambda\beta_{cl}\bar\beta_{cl}e^{\varphi} , \qquad \bar\partial\beta_{cl}=0,\qquad \partial\bar\beta_{cl}=0\nonumber\\
& \bar\partial\gamma_{cl}=\tilde\lambda\bar\beta_{cl} e^{\varphi}+\lambda_0\bar\beta_{cl},\qquad
\partial\bar\gamma_{cl}=\tilde\lambda\beta_{cl} e^{\varphi}+\lambda_0\beta_{cl}
\end{align}
are found to be conserved 
\begin{equation}
\bar\partial T_{cl}(z)=0 \ , \ \ \  \partial \overline{T}_{cl}(\bar z)=0. 
\end{equation}
Notice that this is still true even with the modified equations of motion for $\gamma_{cl}$ and $\bar\gamma_{cl}$ in (\ref{eom}), which involves the term coming from the deformation (\ref{SD}). The gluing conditions in terms of the rescaled fields are 
\begin{equation}
\beta_{cl}+\bar\beta_{cl}|_{z=\bar z}=0 , \ \ \   (\bar\partial-\partial)\varphi|_{z=\bar z}=2\tilde\zeta\beta_{cl}e^{\varphi/2}, \ \ \  \gamma_{cl}+\bar\gamma_{cl}|_{z=\bar z}=\tilde\zeta e^{\varphi/2}+\tilde\lambda_b.
\end{equation}

With the use of the equations of motion and the gluing conditions one can also show that
\begin{equation}
\bar\partial\bar\gamma_{cl}|_{z=\bar z}=
\bar\partial(-\gamma_{cl}+\tilde\zeta e^{\varphi/2}+\tilde\lambda_b)|_{z=\bar z}=
-\tilde\lambda\bar\beta_{cl}e^{\varphi}-\lambda_0\bar\beta_{cl}+\tfrac{1}{2}\tilde\zeta e^{\varphi/2}\partial\varphi+\tilde\zeta^2\beta_{cl}e^{\varphi}
\end{equation}
and
\begin{equation}
\bar\partial\bar\gamma_{cl}|_{z=\bar z}=
-\tilde\lambda\beta_{cl}e^{\varphi}-\lambda_0\beta_{cl}+\tfrac{1}{2}\tilde\zeta e^{\varphi/2}\partial\varphi .
\end{equation}
Therefore, we have
\begin{equation}
-\bar\beta_{cl}\bar\partial\bar\gamma|_{z=\bar z}=-\beta_{cl}\partial\gamma+
\tilde\zeta\beta_{cl} e^{\varphi/2}+\tilde\zeta^2\beta_{cl}^2 e^{\varphi}
\end{equation}
and thus
\begin{equation}
\overline T_{cl}|_{z=\bar z}=-\beta_{cl}\partial\gamma_{cl}+\tilde\zeta\beta_{cl} e^{\varphi/2}\partial\varphi
+\tilde\zeta^2\beta_{cl}^2 e^{\varphi}-\frac{1}{4}\left(\partial\varphi+2\tilde\zeta\beta_{cl}e^{\varphi/2}\right)^2=
T_{cl} \ ,
\end{equation}
which is exactly what we wanted to prove. This justifies the boundary action (\ref{Slab}) as the one preserving (\ref{gluing2}).

\section{Correlation functions and anomalous dimension}

We are interested in computing correlation functions on the disc. We will consider the bulk vertex operator
\begin{equation}
\Phi^{j}({p}|z)=|{p}|^{2(j+1)} e^{{p}\gamma(z)-\bar{p}\bar\gamma(\bar z)}
e^{2b(j+1)\phi(z,\bar z)},\label{La18}
\end{equation}
which is a Kac-Moody primary of the wordsheet CFT. In the undeformed WZW theory ($\lambda_0=0$), this operator has holomorphic and antiholomorphic conformal dimensions $h_j=\bar{h}_j=-b^2 j(j+1)$, where $j$ labels the unitary representation of $SL(2,\mathbb{R})$ to which the state created by (\ref{La18}) belongs. We expect this conformal dimension to receive corrections in the deformed theory, namely to change as
\begin{equation}
h_j\rightarrow h_{\Phi}^{j,{p}}=h_j+\delta h_{\Phi}^{{p}},
\end{equation}
where $\delta h_{\Phi}^{{p}}$ is a $p$-dependent anomalous dimension that vanishes when $\lambda_0=0$. This was studied in \cite{Gaston} by considering the 2-point function on the sphere topology. Here we will consider observables of the deformed theory in the presence of a conformal boundary. On the disc geometry, we will also consider operators of the form
\begin{equation}
\Psi^l(\nu|\tau)=|\nu|^{l+1} e^{\tfrac{1}{2}\nu\gamma(\tau)-\tfrac{1}{2}\nu\bar\gamma(\tau)}
e^{b(l+1)\phi(\tau)},
\end{equation} 
which are inserted at a point $\tau \in \mathbb{R}$ of the boundary $\partial \Gamma $. In the undeformed theory these operators have conformal dimension $h_l=-b^2 l(l+1)$ and, as in the the case of bulk operators, we expect the dimension to be corrected in the deformed theory, namely 
\begin{equation}
h_l\rightarrow h_{\Psi}^{l,\nu}=h_l+\delta h_{\Psi}^{\nu} \ .
\end{equation}

There are three correlation functions whose dependence on the worldsheet coordinates are fully determined by conformal invariance. These are the bulk 1-point function
\begin{equation}\label{1pointB}
\langle \Phi^{j}({p}|z)\rangle_D\sim \frac{1}{|z-\bar z|^{2h_{\Phi}^{j,{p}}}},
\end{equation}
the boundary-boundary 2-point correlation function
\begin{equation}\label{2pointbb}
\langle\Psi^l(\nu|\tau_1)\Psi^l(-\nu|\tau_2)\rangle_D\sim
\frac{1}{|\tau_1-\tau_2|^{2 h_{\Psi}^{l,\nu}}},
\end{equation}
and the bulk-boundary 2-point function
\begin{equation}\label{2pointBb}
\langle \Phi^{j}({p}|z) \Psi^l(\nu|\tau)\rangle_D\sim
\frac{1}{|z-\bar{z}|^{2h_{\Phi}^{j,{p}}-h_{\Psi}^{l,\nu}}|z-\tau|^{2 h_{\Psi}^{l,\nu}}},
\end{equation}
where the subscript $D$ refers to the fact that the expectation values are taken in presence of the deformation.

Our strategy will be as follows: By carefully treating the deformation in the path integral approach of the bulk 1-point function and the boundary-boundary 2-point function, we will obtain two expressions for $\delta h_{\Phi}^{{p}}$ and $\delta h_{\Psi}^{\nu}$, which follow from the expected scalings (\ref{1pointB}) and (\ref{2pointbb}), respectively. Then, with those expressions at hand, we will check the scaling (\ref{2pointBb}) and verify the consistency of our computation, which in particular involves the regularization of logarithmic divergences.

\section{Bulk 1-point function}

As we prove in Appendix A, considering arbitrary values of $\lambda_b$ in $S_b$ does not affect the results as the boundary operator $\int_{\partial \Gamma} dx\beta$ does not contribute to the logarithmic divergence and thus to the anomalous dimension.  So let us set $\lambda_b =0$.

The starting point is then to consider
\begin{equation}
\langle \Phi^j(p|z)\rangle_D \equiv \int {\mathcal D}\beta {\mathcal D}\bar{\beta} {\mathcal D}\gamma {\mathcal D}\bar{\gamma} {\mathcal D}\phi \ e^{-S}\ |{p}|^{2(j+1)} e^{{p}\gamma(z)-\bar{p}\bar\gamma(\bar z)}
e^{2b(j+1)\phi(z,\bar z)}  
\end{equation} 
and to evaluate the path integral following the techniques developed in \cite{HS}. We first integrate out the $\gamma$ and $\bar\gamma$ fields. This yields the Dirac delta
\begin{eqnarray}
\int\mathcal{D}\gamma\, e^{\int \frac{d^2w}{2\pi }\gamma\bar\partial\beta}
e^{{p}\gamma(z)}= 
2\pi \delta\left(\bar\partial\beta(w)-2\pi {p}\delta^{(2)}(w-z)\right)
\end{eqnarray}
and its anti-holomorphic counterpart
\begin{eqnarray}
\int\mathcal{D}\bar\gamma\, e^{\int \frac{d^2w}{2\pi }\bar\gamma\partial\bar\beta}
e^{-\bar{p}\bar\gamma(\bar z)}= 2\pi 
\delta\left(\partial\bar\beta(\bar w)+2\pi \bar{p}\delta^{(2)}(\bar w-\bar z)\right)
\end{eqnarray}

Fields $\beta$ and $\bar{\beta}$ are 1-differentials. The solutions of the two constraints above are compatible with the proper boundary conditions only for ${p}+\bar{p}=0$. They are given by
\begin{equation}\label{betasolutions}
\beta(w)=\frac{p(z-\bar z )}{(w-z)(w-\bar z)},\qquad
\bar\beta(\bar w)=\frac{\bar{p}(z-\bar{z})}{(\bar w-z)(\bar w-\bar z)}
\end{equation}

The rest of the path integral computation parallels exactly \cite{FR, Babaro}, the only difference being that now we have to evaluate the deformation operator $S_D$ on the solution (\ref{betasolutions}). This contributes to the final result with an additional overall factor
\begin{equation}
\langle \Phi^j(p|z)\rangle_D\propto e^{ -\frac{\lambda_0}{\pi} I_{B}(z)},
\end{equation}
where $I_{B}$ is the logarithmically divergent integral
\begin{equation}
I_{B}(z)=|{p}|^2 |z-\bar z|^2\int\limits_{\Gamma} \frac{d^2w}{|w-z|^2|w-\bar{z}|^2}=
\frac{1}{2}|{p}|^2 |z-\bar z|^2\int\limits_{\mathbb{C}} \frac{d^2w}{|w-z|^2|w-\bar{z}|^2}. \label{La56}
\end{equation}
In the second equality we have used the fact that the change $w\leftrightarrow\bar w$ leaves the integrand invariant while mapping the upper half plane into the lower half plane. This means that the original integral is half of the integral in the whole complex plane. 

Since integral (\ref{La56}) is divergent, we need to regularize it in order to extract the logarithmic behavior. We may resort to dimensional regularization: We introduce the regularized version of (\ref{La56}), namely
\begin{equation}
I_{B}^{\epsilon}(z)= \frac 12
{|{p}|^2 |z-\bar z|^2} (l^2e^{\gamma}\pi)^\epsilon\int\limits_{\mathbb{C}} d^{2-2\epsilon}w\frac{1}{|w-z|^{2}|w-\bar{z}|^{2}},
\end{equation}
where we have introduced the scale $l$ and the factor $e^{\gamma \epsilon}\pi^\epsilon$ to absorb irrelevant constants. This integral is easily solved by standard methods, obtaining
\begin{equation}
I_{B}^{\epsilon}(z)=|{p}|^2 l^{2\epsilon}e^{\gamma\epsilon}\pi\frac{\Gamma^2(-\epsilon)\Gamma(1+\epsilon)}{\Gamma(-2\epsilon)|z-\bar z|^{2\epsilon}}=
2|{p}|^2\left(
-\frac{\pi}{\epsilon}+2\pi\log\frac{|z-\bar z|}{l}+\mathcal{O}(\epsilon)
\right).
\end{equation}

Therefore, we obtain that
\begin{equation}
e^{-S_{D}}\simeq \frac{e^{2\lambda_0|{p}|^2/\epsilon}}{|z-\bar z|^{4\lambda_0|{p}|^2}},\label{La59}
\end{equation}
where the symbol $\simeq $ here means that the quantity on the right hand side is what the piece $e^{-S_D}$ of the path integral measure reduces to after evaluation and in the limit $\epsilon \to 0$.

From (\ref{La59}), we can read the correction $\delta h_{\Phi}^{{p}}$ in (\ref{1pointB}), which turns out to be
\begin{equation}
{\delta h_{\Phi}^{{p}}=2\lambda_0|{p}|^2}.
\end{equation}

This means that the conformal dimension of the worldsheet deformed theory is
\begin{equation}
h_{\Phi }^{j,p}=-b^2j(j+1)+2\lambda_0|p|^2.\label{PPP}
\end{equation}
The spectrum of string theory on the interpolating background follows from the Virasoro constraints for (\ref{PPP}). 

The rest of the 1-point function computation goes exactly along the lines of \cite{FR,Babaro}. In other words, the only difference between the computation in the WZW theory and in the deformed theory is expressed by the following relation
\begin{equation}
\langle \Phi^{j}({p}| z)\rangle_{D}=\frac{1}{|z-\bar z|^{4\lambda_0|{p}|^2}}
\langle \Phi^{j}({p}| z)\rangle_{\text{WZW}},\label{subs}
\end{equation}
where a wave function renormalization of the vertex operator 
\begin{equation}\label{renormalizationBulk}
\Phi^{j}({p}| z)\rightarrow \Phi^{j}({p}| z) e^{-2\lambda_0|{p}|^2/\epsilon}
\end{equation}
is needed in order to absorb the pole through the regularization.

Equation (\ref{subs}) gives a closed expression for the 1-point function in the deformed theory in terms of the analogous quantity in the WZW theory. The latter, on the other hand, can be explicitly computed \cite{FR} and thus shown to yield
\begin{equation}
\langle \Phi^{j}({p}| z)\rangle_{D}=|z-\bar{z}|^{-2h_{\Phi}^{j,p}} c_b\ \delta(p+\bar{p})|p|\Gamma(2j+1)\Gamma(1+b^2(2j+1))\cosh (2j+1)
\end{equation}
where $c_b$ is an unimportant ($j$-independent) factor, and where we have fixed $M_0$ to a specific value resorting to the shift symmetry under $\phi \to \phi + \phi_0$.

Before concluding this section, a few words on the regularization scheme are due: Let us go back to integral (\ref{La56}), namely
\begin{equation}
I_{B}(z)=|z-\bar z|^2 |p|^2 \int\limits_{\Gamma} \frac{d^2w}{|w-z|^2|w-\bar{z}|^2}.
\end{equation}
As said, this integral exhibits a logarithmic divergence when $w\to z$. Since we are integrating over half of the complex plane, the point $\bar z$ lies outside the region of integration and therefore it does not produce another divergence. In the computation above we resorted to dimensional regularization. Alternatively, we could have chosen to extract the logarithmic behavior with the tricks employed in \cite{Gaston}, which amounts to consider instead the regularized integral
\begin{equation}
I_{B}^{\epsilon}(z)=
|z-\bar z|^2 |p|^2 \int\limits_{\Gamma} \frac{d^2w}{|w-z|^{2-2\epsilon}|w-\bar{z}|^{2-2\epsilon}}.\label{Mala}
\end{equation}

However, expanding in $\epsilon$ and extracting the $\log|z-\bar z|$ piece, (\ref{Mala}) yields
\begin{equation}
S_{D}\simeq \frac{\lambda_0}{\pi}I_B \simeq 8\lambda_0|{p}|^2\log|z-\bar z| + \ldots
\end{equation}
where the ellipsis stand for contributions other than the logarithmic piece, and therefore
\begin{equation}
e^{-S_{D}}\simeq \frac{1}{|z-\bar z|^{8\lambda_0|{p}|^2}}\ , \label{La42}
\end{equation}
which, after renormalization of the vertex, differs from (\ref{La59}) in a factor 2 in the exponent. This difference is an artifact of the procedure (\ref{Mala}), as we will discuss in detail in Appendix B. In turn, as a byproduct of (\ref{La59}), we correct a factor 2 in the computation of \cite{Gaston}.

\section{Boundary-boundary 2-point functions}

Now, we move to the 2-point function. Consider the correlator of two boundary operators with different momenta $\nu_1$ and $\nu_2$
\begin{equation}
\langle\Psi^l(\nu_1|\tau_1)\Psi^l(\nu_2|\tau_2)\rangle_D \ .
\end{equation}

The path integral over $\gamma$ and $\bar\gamma$ fields now produces the Dirac delta
\begin{eqnarray}
\int\mathcal{D}\gamma\, e^{\int \frac{d^2w}{2\pi}\gamma\bar\partial\beta}
e^{\tfrac{1}{2}\nu_1\gamma(\tau_1)}e^{\tfrac{1}{2}\nu_2\gamma(\tau_2)}= 
2\pi \delta\left(\bar\partial\beta(w)-\pi \sum\limits_{i=1}^{2}\nu_i\delta^{(2)}(w-\tau_i)\right)\nonumber
\end{eqnarray}
and its anti-holomorphic counterpart. The solution exists only for $\nu_1+\nu_2=0$, and is given by
\begin{eqnarray}
\beta(w)&=&\frac{\nu_1}{w-\tau_1}+\frac{\nu_2}{w-\tau_2}=\frac{\nu(\tau_1-\tau_2)}{(w-\tau_1)(w-\tau_2)}\label{Astro}\\
\bar\beta(\bar w)&=&-\frac{\nu_1}{\bar w-\tau_1}-\frac{\nu_2}{\bar w-\tau_2}=\frac{\nu(\tau_2-\tau_1)}{(\bar w-\tau_1)(\bar w-\tau_2)}\label{Boy}
\end{eqnarray}
where we defined $\nu=\nu_1=-\nu_2$. Since $\tau_i$ belongs to the boundary, the factor $\delta^{(2)}(w-\tau_i)$ can be computed by slightly moving the insertions $\tau_i$ inside the bulk and then taking the limit in order to correctly obtain the numerical factors in the solution for $\beta$ and $\bar\beta$. The contribution $S_D$, once evaluated on (\ref{Astro})-(\ref{Boy}), yields
\begin{equation}
S_{D}= \frac{\lambda_0}{\pi} I_{bb}(z)
\end{equation}
with
\begin{equation}
I_{bb}(z)=\nu^2 |\tau_1-\tau_2|^2\int\limits_{\Gamma} \frac{d^2w}{|w-\tau_1|^2|w-\tau_2|^2}=\frac 12 
{\nu^2 |\tau_1-\tau_2|^2}\int\limits_{\mathbb{C}} \frac{d^2w}{|w-\tau_1|^2|w-\tau_2|^2},
\end{equation}
where, again, in the second equality we halved the result by extending the integral to the whole complex plane. 
$I_{bb}$ is also divergent; its regularized version would be 
\begin{equation}
I_{bb}^{\epsilon}(z)=\frac 12 
{\nu^2 |\tau_1-\tau_2|^2} (l^2e^{\gamma}\pi)^\epsilon\int\limits_{\mathbb{C}} \frac{d^{2-2\epsilon}w}{|w-\tau_1|^2|w-\tau_2|^2},
\end{equation}
which is completely analogous to the integral of the previous section. In fact, we get
\begin{equation}
I_{bb}^{\epsilon}(z)=
2\nu^2\left(
-\frac{\pi}{\epsilon}+2\pi\log\frac{|\tau_1-\tau_2|}{l}+\mathcal{O}(\epsilon)
\right)
\end{equation}
and, finally, the contribution of the deformation operator to the path integral gives
\begin{equation}
e^{-S_{D}}\simeq \frac{e^{2\lambda_0\nu^2/\epsilon}}{|\tau_1-\tau_2|^{4\lambda_0\nu^2}}\label{La47}
\end{equation}
from which we read the correction $\delta h_{\Psi}^{\nu}$ using (\ref{2pointbb}); namely
\begin{equation}
{\delta h_{\Psi}^{\nu}=2\lambda_0\nu^2},
\end{equation}
which determines the spectrum of the boundary operators in the worldsheet theory. 

Eventually, we find
\begin{equation}
\langle\Psi^l(\nu|\tau_1)\Psi^l(-\nu|\tau_2)\rangle_{D}=\frac{1}{|\tau_1-\tau_2|^{4\lambda_0\nu^2}}
\langle\Psi^l(\nu|\tau_1)\Psi^l(-\nu|\tau_2)\rangle_{\text{WZW}},
\end{equation}
where, as in the case of the bulk 1-point function, the vertex operators $\Psi^l(\nu|\tau)$ need to be renormalized by a factor $e^{-{\lambda_0\nu^2}/{\epsilon}}$. 

\section{Bulk-boundary 2-point functions}

A non-trivial consistency check of the results obtained in the previous sections follows from the computation of the boundary-bulk correlator $\langle \Phi^{j}({p}|z) \Psi^l(\nu|\tau)\rangle_D$. As in the previous cases, after integrating over $\gamma$ and $\bar\gamma$ fields, we obtain a pair of Delta functions whose solutions exist for ${p}+\bar{p}+\nu=0$. They are given by
\begin{equation}
\beta(w)=\frac{{p}}{w-z}+\frac{\bar{p}}{w-\bar z}+\frac{\nu}{w-\tau},\quad
\bar\beta(\bar w)=-\frac{{p}}{\bar w-z}-\frac{\bar{p}}{\bar w-\bar z}-\frac{\nu}{\bar w-\tau}.
\end{equation}

Using $\nu=-{p}-\bar{p}$ we may regroup the denominators
\begin{align}
& \beta(w)=\frac
{{p}(w-\bar z)(z-\tau)+\bar{p}(w-z)(\bar z-\tau)}
{(w-z)(w-\bar z)(w-\tau)},\nonumber\\
& \bar\beta(\bar w)=-\frac
{{p}(\bar w-\bar z)(z-\tau)+\bar{p}(\bar w-z)(\bar z-\tau)}
{(\bar w-z)(\bar w-\bar z)(\bar w-\tau)}.
\end{align}

Evaluating these solutions on the deformation operator $S_D$, we obtain
\begin{equation}\label{defIBb}
S_{D}=-\frac{\lambda_0}{\pi}\int\limits_{\Gamma}d^2 w\beta(w)\bar\beta(w)\rightarrow
\frac{\lambda_0}{\pi}\left(I_{Bb}^{(1)}(z,\tau)+I_{Bb}^{(2)}(z,\tau)+I_{Bb}^{(3)}(z,\tau)\right)
\end{equation}
where we define the (still unregularized) integrals
\begin{align}
 I_{Bb}^{(1)}(z,\tau)  =\frac 12  {|{p}|^2|z-\tau|^2}\int\limits_{\mathbb{C}} d^2w
\left(\frac{1}{|w-z|^2|w-\tau|^2}+\frac{1}{|w-\bar z|^2|w-\tau|^2}\right)\label{IBb1},
\end{align}
\begin{align}
 I_{Bb}^{(2)}(z,\tau) =\frac 12 {\bar{p}^2(\bar z-\tau)^2}\int\limits_{\mathbb{C}} d^2w
\frac{(w-z)(\bar w-z)}{|w-z|^2|w-\bar z|^2|w-\tau|^2}\label{IBb2},
\end{align}
and
\begin{align}
 I_{Bb}^{(3)}(z,\tau)= \frac 12 {{p}^2( z-\tau)^2}\int\limits_{\mathbb{C}} d^2w
\frac{(\bar w-\bar z)( w-\bar z)}{|w-z|^2|w-\bar z|^2|w-\tau|^2},\label{IBb3}
\end{align}
where $ I_{Bb}^{(3)}(z,\tau)= (I_{Bb}^{(2)}(z,\tau))^{*}$.

Integral $I_{Bb}^{(1)}(z,\tau)$ is completely analogous to the integrals we regularized and calculated in the two previous sections. Its regularized version $I_{Bb}^{(1,\epsilon)}(z,\tau)$ results in
\begin{equation}\label{IBb1r}
I_{Bb}^{(1,\epsilon)}(z,\tau)=4\pi|{p}|^2\left(-\frac{1}{\epsilon}+2\log\frac{|z-\tau|}{l}+\mathcal{O}(\epsilon)\right).
\end{equation}

In contrast, integrals $I_{Bb}^{(2)}(z,\tau)$ and $I_{Bb}^{(3)}(z,\tau)$ are much more involved and are solved in Appendix \ref{Integrals}. Here we just write down their results
\begin{align}\label{IBb23r}
I_{Bb}^{(2,\epsilon)}(z,\tau)= & \pi\bar{p}^2\left(
-\frac{1}{\epsilon}-2\log\frac{|z-\bar z|}{l}+4\log\frac{|z-\tau|}{l}+\mathcal{O}(\epsilon)
\right),\nonumber\\
I_{Bb}^{(3,\epsilon)}(z,\tau)= & \pi{p}^2\left(
-\frac{1}{\epsilon}-2\log\frac{|z-\bar z|}{l}+4\log\frac{|z-\tau|}{l}+\mathcal{O}(\epsilon)
\right).
\end{align}

Using (\ref{IBb1r}) and (\ref{IBb23r}) in (\ref{defIBb}), we get
\begin{equation}
S_{D}\simeq
-\frac{\lambda_0}{\epsilon}\left(2|{p}|^2+\nu^2\right)
+\lambda_0\left(
(4|{p}|^2-2\nu^2)\log\frac{|z-\bar z|}{l}+4\nu^2 \log\frac{|z-\tau|}{l}
\right)+\mathcal{O}(\epsilon),
\end{equation}
where we used the simple property that since $\nu=-{p}-\bar{p}$, we have ${p}^2+\bar{p}^2=\nu^2-2|{p}|^2$. Finally, the deformation operator contributes to the path integral with
\begin{equation}\label{BbExponential}
e^{-S_{D}}\simeq
\frac{e^{\tfrac{2\lambda_0}{\epsilon}|{p}|^2+\tfrac{\lambda_0}{\epsilon}\nu^2}}
{|z-\bar z|^{4\lambda_0|{p}|^2-2\lambda_0\nu^2}|z-\tau|^{4\lambda_0\nu^2}},
\end{equation}
which is exactly the power dependence on $|z-\bar z|$ and $|z-\tau|$ we expected (cf.\ (\ref{2pointBb})), showing the consistency with our previous computations of the anomalous dimensions $\delta h_{\Phi}^{{p}}=2\lambda_0|{p}|^2$ and $\delta h_{\Psi}^{\nu}=2\lambda_0\nu^2$. Moreover, the renormalization of the operators we had proposed before, namely
\begin{equation}
\Phi^{j}({p}| z)\rightarrow \Phi^{j}({p}| z) e^{-\tfrac{2\lambda_0|{p}|^2}{\epsilon}}
,\qquad \Psi^l(\nu|\tau)\rightarrow \Psi^l(\nu|\tau) e^{-\tfrac{\lambda_0\nu^2}{\epsilon}}
\end{equation}
exactly cancels the poles in (\ref{BbExponential}) allowing us to drop the regulator. In conclusion, the correlator computation leads to the relation
\begin{equation}
\langle \Phi^{j}({p}|z) \Psi^l(\nu|\tau)\rangle_{D}=
\frac{1}
{|z-\bar{z}|^{2\delta h_{\Phi}^{{p}}-\delta h_{\Psi}^{\nu}}
|z-\tau|^{2 \delta h_{\Psi}^{\nu}}}
\langle \Phi^{j}({p}|z) \Psi^l(\nu|\tau)\rangle_{\text{WZW}}
\end{equation}
with exactly $\delta h_{\Phi}^{{p}}$ and $\delta h_{\Psi}^{\nu}$ that we obtained before.

\section{Perturbation theory}

As a further consistency check of our results, in this section we show how the perturbative approach, based on the Coulomb gas realization of the worldsheet correlation functions, reproduces the path integral results obtained in sections 5 and 6.

The Coulomb gas realization amounts to considering the free field theory perturbed by the bulk operator
\begin{equation}
\frac{1}{2\pi }\int_{\Gamma } d^2 z \beta \bar{\beta} (b^2M_0e^{2b\phi}+ 2\lambda_0 ) 
\end{equation} 
and the boundary operator 
\begin{equation}
\frac{i}{4\pi}\int_{\partial \Gamma } dx \zeta \beta\ e^{b\phi},
\end{equation} 
which will appear in the expectation values as integrated screening charges. The number of such operators present in the correlators depends on the momenta of the external states and is determined by the integration over the zero-mode of the free fields; see (\ref{res}) below.

The gluing conditions for the free theory are given by
\begin{equation}\label{glue}
	\beta + \bbeta |_{z=\bar z} = 0 \ , \ \ \ \gamma + \bgamma |_{z=\bar z}= 0 \ , \ \ \ 	(\partial - \bpartial)\phi |_{z=\bar z} = 0\ ,
\end{equation}
cf.\ (\ref{gluing1}). The non-vanishing expectation values of the fields in the presence of the gluing conditions \eqref{glue} are
\begin{eqnarray}\label{pairings0}
	\langle \phi(z)\phi(w)\rangle &= -\log|z-w||\bar{z} - w|
\end{eqnarray}
and
\begin{align}\label{pairings}
\begin{split}
	\langle\beta(z)\gamma(w)\rangle &= \frac{1}{w-z}\ , \ \ \ \ \langle\bbeta(\bar{z})\gamma(w)\rangle = \frac{1}{\bar{z}-w}\\
	\langle\beta(z)\bgamma(\bar{w}\rangle &= \frac{1}{z-\bar{w}}\ , \ \ \ \ \langle\bbeta(\bar{z})\bgamma(\bar{w})\rangle = \frac{1}{\bar{w}-\bar{z}}.
\end{split}
\end{align}

Following standard techniques \cite{FR}, we obtain an expression for the residue of the resonant 1-point function, namely
\begin{align}\label{res}
\begin{split}
	&\underset{2j+1=-n}{\operatorname{Res}} \langle \Phi^j({p}|z)\rangle_D = \frac{1}{2b} |{p}|^{2j+2} \sum_{m,l,t=0}^{\infty} \delta_{2m+l,n} \frac{1}{m!l!t!} \prod_{i=1}^{m} \int_{\Gamma} d^2 w_i \prod_{k=1}^{k} \int_{\partial \Gamma} d x_k \prod_{r=1}^{t} \int_{\Gamma } d^2 q_r\\
	&\Big\langle e^{{p}\gamma(z)- \bar{p}\bgamma(\bar{z})} e^{2b(j+1)\phi(z,\bar{z})}
	\prod_{i=1}^{m} \frac{M_0 b^2}{2\pi} \beta(w_i) \bbeta(\bar{w}_i) e^{2b\phi(w_i,\bar{w}_i)}
	\prod_{k=1}^{\ell} \frac{i \zeta }{4\pi} \beta(x_k) e^{b \phi(x_k)}
	\prod_{r=1}^{t} \frac{\lambda_0}{\pi} \beta(q_r) \bbeta(\bar{q}_r)
	\Big\rangle ,
\end{split}
\end{align}
where $2j+1=-n$ with $n\in \mathbb{Z}_{\geq 0}$. The integrations on $w_i$ and $ q_r$ are performed over the upper half plane, while the integration over $x_k$ is along the real line. In the following, we omit writing out the symbol ${\operatorname{Res}}$, which is implicit.

The expectation value is to be computed in the free theory. Let us choose the location of the operator on the imaginary axis, say $z=iy$. Then, we obtain the following contributions
\begin{align}
X &= \Big\langle e^{2b(j+1)\phi(iy)} \prod_{i=1}^m e^{2b\phi(w_i)}  \prod_{k=1}^\ell e^{b\phi(x_k)} \Big\rangle
= {|2y|^{-\frac{b^2}{2}(n-1)^2}}
\left( \prod_{k=1}^\ell (y^2+x_k^2) \prod_{i=1}^m|y^2+w_i^2|^2\right)^{b^2(n-1)}  \cdot \nonumber
\\ &\ \ \ \ \cdot 
\left(\prod_{i=1}^{m}\prod_{k=1}^{\ell}|w_i-x_k|^2\prod_{i<i'}^m|w_i-w_{i'}|^2\prod_{i=1}^m\prod_{i'=1}^m|w_i-\bar{w}_{i'}| \prod_{k<k'}^{\ell}|x_k-x_{k'}|\right)^{-2b^2}\\
	Y &= \Big\langle e^{{p}\gamma(iy)-\bar{{p}}\bgamma(-iy)} \prod_{i=1}^m \frac{b^2M_0}{4\pi}\beta(w_i) \bbeta(\bar{w}_i) \prod_{k=1}^\ell \frac{i\zeta}{2\pi}\beta(x_k)\Big\rangle = 2\pi \delta({p}+\bar{{p}}) \left(-\frac{M_0b^2}{2\pi}\right)^m |2y{p}|^n\cdot \nonumber\\
	&\ \ \ \ \cdot \left(-\frac{i\zeta }{4\pi} \right)^{\ell} \prod_{i=1}^m\frac{1}{|y^2+w_i^2|^2} \prod_{k=1}^\ell \frac{1}{(y^2+x_k^2)}\\
	Z &= \Big\langle e^{{p}\gamma(iy)-\bar{{p}}\bgamma(-iy)} \prod_{r=1}^{t} \frac{\lambda_0}{\pi} \beta(q_r) \bbeta(\bar{q}_r)\Big\rangle= \left(-\frac{\lambda_0}{\pi}\right)^t |2y{p}|^{2t} \prod_{r=1}^{t} \frac{1}{|y^2+q_r^2|^2},
\end{align}
which follows from (\ref{pairings0}) and (\ref{pairings}). We are assuming here that the imaginary part of $p$ is positive; otherwise, $\zeta $ changes its sign in the expressions above. 

Notice that $Z$ does not depend on $w_i$ nor on $x_k$. After elementary rearrangement, we can write (\ref{res}) in the following form
\begin{align}
	&\langle \Phi^j({p}|z)\rangle_D = \left(
		\frac{1}{2b} |{p}|^{2j+2} \sum_{m,l=0}^{\infty} \delta_{2m+l,n} \frac{1}{m!l!} \prod_{i=1}^{m} \int_{\Gamma } d^2 w_i \prod_{k=1}^{k} \int_{\partial \Gamma} d x_k \ X \cdot Y \right)
		\cdot \sum_{t=0}^{\infty} \frac{1}{t!} \prod_{r=1}^{t} \int_{\Gamma } d^2 q_r \ Z .
\end{align}

From this, we observe that the expression is a product of the result for the unperturbed WZW expectation value obtained in \cite{FR} times the new factor
\begin{equation}
	\sum_{t=0}^{\infty} \frac{1}{t!} \prod_{r=1}^{t} \int_{\Gamma } d^2 q_r \ Z =  \sum_{t=0}^{\infty} \frac{1}{t!} \left(-\frac{\lambda_0}{\pi} |2y{p}|^{2} \int_{\Gamma} \frac{d^2 q}{|y^2+q^2|^2}\right)^t= e^{-\frac{\lambda_0}{\pi}I_B(iy)},
\end{equation}
where $I_B$ is exactly the integral (\ref{La56}) obtained before. That is, the Coulomb gas computation confirms our path integral computation of the 1-point function. 

For the 2-point function, the Coulomb gas approach also yields the correct result. To see this, in the case of the boundary-boundary 2-point function, for example, one can use the formula
\begin{eqnarray}
\Big\langle e^{\frac 12 \nu \gamma(z_1)} e^{-\frac 12 \nu \gamma(z_2)} \prod_{r=1}^n \beta(w_r) \Big\rangle = \Big(\frac {\nu}{2}\Big)^n{(z_1-z_2)^{n}}{\prod_{r=1}^n(w_r-z_1)^{-1}(w_r-z_2)^{-1}}
\end{eqnarray}
and verify that it leads to reproduce (\ref{La47}) in perfect agreement. This is a further crosscheck of our results for the anomalous dimensions. 

\[
\]

The work of G.G. is supported in part by the NSF through grant PHY-1214302. The work of M.L. is supported in part by ANPCyT (Argentina) through grant PICT-2015-1633.

\appendix

\section{Boundary integrals}\label{Appa}

In this appendix, we justify the choice $\lambda_b =0$ in the computation performed in section 4. More precisely, we show that the choice $\lambda_b \neq 0$ would not affect the result for the anomalous dimension. To do so, we go back to the boundary action
\begin{equation}
S_{{b}}\ =\ \frac{1}{2\pi }\int_{\partial \Gamma
}dx\,g^{1/4}\left( bK\phi \,+\,\frac{i\beta}{2} \left( \gamma +\bar{\gamma}\right)-\frac{i\zeta}{2} \,\beta\,e^{b\phi }\,-\,\frac{i\lambda_{b}}{2}\beta \right) ,
\label{conboundaries}
\end{equation}%
with arbitrary $\zeta $ and $\lambda_{b}$, and we will prove that the fourth term does not contribute to the prefactor of the logarithmic divergence. We recall the conventions: $z=x+iy, \bar{z}=x-iy$ and $d^2z = 2dxdy$, so that
\begin{equation}
\partial = \frac{\partial}{\partial z} =\frac{1}{2}\partial_x-\frac{i}{2}\partial_y \ , \ \ \ \bar{\partial} = \frac{\partial}{\partial \bar{z}} = \frac{1}{2}\partial_x+\frac{i}{2}\partial_y.
\end{equation}
In particular, this yields
\begin{equation}
\int_{\mathbb{R}_{\geq 0}} dy\ \partial_y(\beta\gamma)=\int_{\mathbb{R}_{\geq 0}}dy\ \left(\partial_y(\bar{\beta}\bar{\gamma})\right)=-i\beta\gamma\big\vert_{y=0}.
\end{equation}

This is used to show that the total action takes the form we discussed before, namely
\begin{equation}
\begin{split}
S\ &=\ \frac{1}{2\pi }\int_{\Gamma }d^{2}z\,g^{1/2}\left( \partial \phi
\bar{\partial}\phi -\gamma \bar{\partial}\beta -\bar{\gamma}\partial \bar{%
\beta}+\frac{b}{4}R\phi - b^2M_0\beta \bar{\beta}\,e^{2b\phi} -2 \lambda_0\beta\bar{\beta}
\right) + \\
&\ \   \ \ \frac{1}{4\pi }\int_{\partial \Gamma }dx\,g^{1/4}\left( 2bK\phi -i\zeta\beta\,e^{b\phi} -i\lambda_b\beta \right)
\end{split}
\label{S}
\end{equation}

Path integration over fields $\gamma$ and $\bar{\gamma}$ compatible with the boundary conditions $\beta +\bar{\beta}_{|z=\bar{z}}=0$ yields the solutions
\begin{eqnarray}
\beta (w) =  \frac{{p}(\bar{z}-z)}{(w-z)(w-\bar{z})},  \ \ \ \bar{\beta}(\bar{w})= \frac{\bar{{p}}\,(\bar{z}-z)}{(\bar{w}-z)(\bar{w}-\bar{z})}.
\label{cuatro}
\end{eqnarray}

Following \cite{HS} closely, we evaluate the full action on the solutions (\ref{cuatro}) for $\beta $ and $\bar{\beta}$, what results in 
\begin{equation}
\begin{split}
\left\langle \Phi ^{j}({p} |z)\right\rangle _{D} &=\int \mathcal{D}\phi \ \exp\left(- \int_{\Gamma} \frac{d^2w}{2\pi }\left( \partial \phi \bar{\partial}\phi\,+\,\frac{b}{4}R\phi\,+\,\frac{|p|^2|z-\bar{z}|^2}{|w-z|^2|w-\bar{z}|^2}(b^2M_0e^{2b\phi}+2\lambda_0)\right) \right) \cdot \\
& \ \ \ \  \ \ \ \ \cdot \exp \left( -\int_{\partial\Gamma } \frac{dx}{4\pi } \left(2b K\phi\,-\frac{ip(\bar{z}-z)}{(x-z)(x-\bar{z})}(\zeta e^{b\phi}+\lambda_b) \right)\right)\cdot \\
& \ \ \ \  \ \ \ \ \ \cdot |{p} |^{2(j+1)}\ e^{2b(j+1)\phi (z,\bar{z})}\delta
^{(2)}({p} +\bar{{p}}).
\end{split}%
\end{equation}

This reduces the computation of the 1-point function to a Liouville theory computation \cite{FZZ} times a prefactor. Such prefactor differs from the one in the unperturbed theory by two contributions. These are
\begin{equation}
e^{-\frac{\lambda_0}{\pi}I_B(z)} \ \ \ \ \ \text{with} \ \ \ \ \  
I_B = \frac{1}{2}|p|^2 |z-\bar{z}|^2\int_{\mathbb{C}} \frac{d^2w}{|w-z|^2|w-\bar{z}|^2},\label{La877}
\end{equation}
with which we dealt in section 5, and
\begin{equation}
e^{-i\frac{\lambda_b}{\pi }I_b(z)} \ \ \ \ \ \text{with}\ \ \ \ \ I_b = p(z-\bar{z})\int_{\mathbb{R}} \frac{dx}{(x-z)(x-\bar{z})} = 2\pi i p .\label{La878}
\end{equation}

Unlike (\ref{La877}), integral (\ref{La878}) is finite, as it can be easily verified by evaluating the residue of the integrand on $\Gamma $. Therefore, we conclude that the boundary operator $\int_{\partial \Gamma} dx\beta$ does not contribute to the logarithmic divergence, and this justifies setting $\lambda_b =0$ in the computation of the anomalous dimensions in section 5. 


\section{Regularization schemes}\label{AppB}

In this appendix, we discuss in detail different regularization schemes to solve the logarithmically divergent integrals we have been involved with. Let us go back to integral (\ref{La56}), namely
\begin{equation}
I_{B}(z)=|z-\bar z|^2 |p|^2 \int\limits_{\Gamma} \frac{d^2w}{|w-z|^2|w-\bar{z}|^2}.
\end{equation}

This integral has a logarithmic divergence when $w\to z$. As mentioned before, the point $w=\bar z$ lies outside the region of integration and it does not produce another divergence.

The question is, what is the efficient way of dealing with the divergence in (\ref{La877})? Let us begin by reviewing the regularization method employed in \cite{Gaston}, which amounts to introducing the regularized version of the integral
\begin{equation}
I_{B}^{\epsilon}(z)=
|z-\bar z|^2 |p|^2 \int\limits_{\Gamma} \frac{d^2w}{|w-z|^{2-2\epsilon}|w-\bar{z}|^{2-2\epsilon}}.
\end{equation}
We can write this integral using real coordinates: we call $z=x+iy$ and $w=w_1+iw_2$. We then have
\begin{equation}
I_{B}^{\epsilon}(z)=
|z-\bar z|^2 |p|^2 \int_{\mathbb{R}}dw_1
\int_{\mathbb{R}_{>0}}dw_2 
\frac{1}{\left[(w_1-x)^2+(w_2-y)^2\right]^{1-\epsilon}
\left[(w_1-x)^2+(w_2+y)^2\right]^{1-\epsilon}}.
\end{equation}
The trivial change of variables $w_2\to-w_2$ leaves the integrand invariant, namely
\begin{equation}
I_{B}^{\epsilon}(z)=
|z-\bar z|^2 |p|^2 \int_{\mathbb{R}}dw_1
\int_{\mathbb{R}_{<0}}dw_2 
\frac{1}{\left[(w_1-x)^2+(w_2+y)^2\right]^{1-\epsilon}
\left[(w_1-x)^2+(w_2-y)^2\right]^{1-\epsilon}}.
\end{equation}
Thus, integrating in the upper half plane is the same as integrating in the lower half plane. Therefore, the integral we are aiming for is half the integral in the whole complex plane
\begin{equation}
I_{B}^{\epsilon}(z)=\frac{ 1}{2}
|z-\bar z|^2  |p|^2  \int\limits_{\mathbb{C}} \frac{d^2w}{|w-z|^{2-2\epsilon}|w-\bar{z}|^{2-2\epsilon}}.
\end{equation}
This is a Shapiro-Virasoro integral. Integrating it, we obtain
\begin{equation}
I_{B}^{\epsilon}(z)=\pi  |p|^2 |z-\bar z|^{4\epsilon} 
\frac{\Gamma^2(\epsilon)\Gamma(1-2\epsilon)}{\Gamma(2\epsilon)\Gamma^2(1-\epsilon)}.
\end{equation}

Finally, expanding in $\epsilon$ and extracting the $\log|z-\bar z|$ piece we obtain
\begin{equation}
-S_{D}\simeq -8\lambda_0|{p}|^2\log|z-\bar z|+\ldots\label{La106}
\end{equation}
and therefore
\begin{equation}
e^{-S_{D}}\simeq \frac{e^{2\lambda_0 |p|^2/\epsilon}}{|z-\bar z|^{8\lambda_0|{p}|^2}},\label{La107}
\end{equation}
which, as we mentioned in section 5, differs from the dimensional regularization result (\ref{La59}) in a factor 2 in the exponent. This does not change the physics of the problem, as the precise value of $\lambda_0$ can be changed by shifting the zero-mode of the linear dilaton \cite{Gaston}. However, it is still worthwhile understanding the origin of the discrepancy in a factor 2 between (\ref{La107}) and the dimensional regularization result (\ref{La59}). We will argue that the latter gives the correct value, which we will confirm below by three different methods. 

Dimensional regularization amounts to replacing
\begin{equation}
S_{D}=-\frac{\lambda_0}{\pi}\int\limits_{\Gamma}d^2z\,g^{1/2}\beta\bar\beta \ 
\rightarrow \ 
-\frac{\lambda_0\, l_0^{2\epsilon}}{\pi}\int\limits_{\Gamma}d^{2-2\epsilon}z\,g^{1/2}\beta\bar\beta,
\end{equation}
where a scale $l_0$ is introduced. This leads to the regularized integral
\begin{equation}
I_{B, \epsilon}^{(1)}=\frac 12 l_0^{2\epsilon}|z_1-z_2|^2  |p|^2 \int\limits_{\mathbb{C}}
\frac{d^{2-2\epsilon}z}{|z-z_1|^{2}|z-z_2|^{2}}
\end{equation}
with solution
\begin{equation}
I_{B, \epsilon}^{(1)}=l_0^{2\epsilon} |p|^2 |z_1-z_2|^{-2\epsilon}\pi^{1-\epsilon}
\frac{\Gamma^2(-\epsilon)\Gamma(1+\epsilon)}{\Gamma(-2\epsilon)}.
\end{equation}

Expanding in $\epsilon$ and extracting the logarithm we get
\begin{equation}
S_D= \frac{\lambda_0}{\pi}I^{(1)}_B\simeq 4 |p|^2 \lambda_0 \log\frac{|z_1-z_2|}{l_0} + \ldots
\end{equation}
which is half of (\ref{La106}), up to a constant term.

Another way of introducing a consistent regularization, somehow closer to the one used in \cite{Gaston}, would be to slightly change the power of the $\beta\bar\beta$ term in the deformation operator $S_D$. This would, once again, take the theory slightly away from marginality and introduce a natural way of regularizing the integrals. That is, one replaces
\begin{equation}
S_{D}=-\frac{\lambda_0}{\pi}\int\limits_{\Gamma}d^2z\,g^{1/2}\beta\bar\beta \ 
\rightarrow \ 
-\frac{\lambda_0\, l_0^{-2\epsilon}}{\pi}\int\limits_{\Gamma}d^{2}z\,g^{1/2}
(\beta\bar\beta)^{1-\epsilon},
\end{equation}
where, again, a scale $l_0$ is introduced. When solving the functional integral and making the substitution of the fields $\beta$, we obtain the integral
\begin{equation}
I_{B, \epsilon}^{(2)}=\frac 12 l_0^{-2\epsilon}|z_1-z_2|^{2-2\epsilon} |p|^2 \int\limits_{\mathbb{C}}\frac{d^2z}{|z-z_1|^{2-2\epsilon}|z-z_2|^{2-2\epsilon}}.
\end{equation}
Notice that this regularized integral is almost the same as the one in \cite{Gaston} except for the modified power of $|z_1-z_2|$ in front of the integral. Solving it, we get
\begin{equation}
I_{B, \epsilon}^{(2)}=\pi |p|^2 \frac{|z_1-z_2|^{2\epsilon}}{l_0^{2\epsilon}}
\frac{\Gamma^2(\epsilon)\Gamma(1-2\epsilon)}{\Gamma(2\epsilon)\Gamma^2(1-\epsilon)}
\end{equation}
and expanding in $\epsilon$ and extracting the logarithm we get
\begin{equation}
S_D = \frac{\lambda_0}{\pi }I^{(2)}_B\simeq 4 |p|^2 \lambda_0 \log\frac{|z_1-z_2|}{l_0} + \ldots
\end{equation}
which is again half of (\ref{La106}).

A third way of obtaining the same result --less systematic but still widely used in the context of extracting logarithmic divergences in spacetime integrals for anomalous dimensions-- is the following: Consider now the integral
\begin{equation}
I_{B, l_0}^{(3)}=\frac 12 |z_1-z_2|^2 |p|^2 \int\limits_{\mathbb{C}\backslash\{z_1,z_2\}_{l_0}}\!\!
\ \frac{d^2z}{|z-z_1|^2|z-z_2|^2}
\end{equation}
where we introduce, as a regulator, the fact that we integrate in the whole complex plane except for two small circles of radius $l_0$ centered at $z_1$ and $z_2$. It is clear that the logarithmic divergences will appear when integrating in the region close to $z_1$ and $z_2$. Therefore, we separate the integral in three regions: two annular regions around the singularities $z_1$ and $z_2$, and the rest of the complex plane.

Consider first the annulus around $z_1$. The smaller radius would be the cutoff $l_0$ and one would need to define the bigger radius. Since we cannot integrate further than the position of $z_2$, in order not to overlap integrals, the biggest radius should be $|z_1-z_2|/2$. Before writing this down, notice that this contribution will be equivalent to the second annular region and therefore we just multiply the contribution by 2. Using the parametrization $z=z_1+r e^{i\theta}$ we get
\begin{equation}
I_{B,l_0}^{(3)}=2|z_1-z_2|^2 |p|^2 \int\limits_{l_0}^{|z_1-z_2|/2}dr\int\limits_0^{2\pi}d\theta
\frac{r}{r^2\,|z_1-z_2+r e^{i\theta}|^2}+\ldots ,
\end{equation}
where the ellipsis stand for integration in the regions which do not contribute to log divergences. Since we are only interested in the integration for small $r$, where the measure $dr/r$ is divergent, we may approximate $|z_1-z_2+r e^{i\theta}|^2$ by $ |z_1-z_2|^2$. Integrating, we get
\begin{equation}
S_D = \frac {\lambda_0 }{\pi} I_{B}^{(3)}\simeq 4  |p|^2 \lambda_0 \log\frac{|z_1-z_2|}{\tilde{l}_0}+\ldots,
\end{equation}
which, again, is one half of (\ref{La106}), in perfect agreement with (\ref{La59}).

\section{Conformal integrals for the 2-point function}\label{Integrals}

We had postponed the computation of the integral
\begin{equation}
I_{Bb}^{(2)}(z,\tau) = \frac{\bar{p}^2(\bar z-\tau)^2}{2}\int\limits_{\mathbb{C}} d^2w
\frac{(w-z)(\bar w-z)}{|w-z|^2|w-\bar z|^2|w-\tau|^2}
\end{equation}
and $I_{Bb}^{(3)}(z,\tau)$, which are related by $I_{Bb}^{(3)}(z,\tau)=(I_{Bb}^{(2)}(z,\tau))^*$.
We compute this integral here: Let us first separate the numerator with the obvious property
\begin{equation}
(w-z)(\bar w-z)=|w-z|^2+(w-z)(\bar z-z),
\end{equation}
effectively obtaining
\begin{equation}\label{separationIntegrals}
I_{Bb}^{(2)}(z,\tau) = \frac{\bar{p}^2(\bar z-\tau)^2}{2}
\left(\mathcal{B}(|z-\tau|^2)+(\bar z-z)\mathfrak{T}(z,\tau)\right)
\end{equation}
where $\mathcal{B}(|z-\tau|^2)$ is the usual bubble integral which we already know how to regularize
\begin{equation}\label{Bubble}
\mathcal{B}_{\epsilon}(|z-\tau|^2)=(l^2 e^{\gamma}\pi)^{\epsilon}
\int\limits_{\mathbb{C}} 
\frac{d^{2-2\epsilon}w}{|w-\bar z|^2|w-\tau|^2}=
\frac{4\pi}{|z-\tau|^2}\left(-\frac{1}{\epsilon}+2\log\frac{|z-\tau|}{l}+
\mathcal{O}(\epsilon)\right),
\end{equation}
while $\mathfrak{T}(z,\tau)$ is the principal problem we want to solve in this appendix
\begin{equation}
\mathfrak{T}(z,\tau)=
\int\limits_{\mathbb{C}} d^2w
\frac{(w-z)}{|w-z|^2|w-\bar z|^2|w-\tau|^2}.
\end{equation}
To study it, let us first define the ``star'' (regularized) D-dimensional vector integral
\begin{equation}
\mathcal{T}_{\epsilon}^{\alpha}(x_1,x_2,x_3)=(l^2 e^{\gamma}\pi)^{\epsilon}\int d^D x_0\frac{(x_0-x_1)^{\alpha}}{|x_0-x_1|^2|x_0-x_2|^2|x_0-x_3|^2}.
\end{equation}
In $D=2-2\epsilon$, the vectors $x_0,\ldots,x_4$ have $D$ components that reduce to only 2 components in the limit of $\epsilon\to 0$. Therefore, in this limit, we can associate the two components of those vectors with the real and imaginary parts of our complex plane points $w$, $z$, $\bar z$ and $\tau$. More precisely, we associate
\begin{equation}\label{association}
x_0\to w,\quad x_1\to z,\quad x_2\to \bar z,\quad x_3\to\tau.
\end{equation}
Therefore, if we are able to compute $\mathcal{T}_{\epsilon}^{\alpha}(x_1,x_2,x_3)$ and expand it close to $\epsilon=0$, we can associate the two components of the $\mathcal{T}_{\epsilon}^{\alpha}$ vector with the real and imaginary parts of the regularized version of the integral $\mathfrak{T}(z,\tau)$ we are trying to perform. Thus, in the same sense of the association (\ref{association}) we have that 
\begin{equation}
\mathcal{T}_{\epsilon}^{\alpha}(x_1,x_2,x_3)\to\mathfrak{T}_{\epsilon}(z,\tau),
\end{equation}
or more explicitly $\text{Re}(\mathfrak{T}_{\epsilon}(z,\tau))=\mathcal{T}_{\epsilon}^{\alpha=1}$ and $\text{Im}(\mathfrak{T}_{\epsilon}(z,\tau))=\mathcal{T}_{\epsilon}^{\alpha=2}$.
To solve $\mathcal{T}_{\epsilon}^{\alpha}$ we start with the Passarino-Veltman method. Since it is a translationally invariant vector integral, it can only be proportional to difference vectors
\begin{equation}\label{PVansatz}
\mathcal{T}_{\epsilon}^{\alpha}(x_1,x_2,x_3)=A\, x_{21}^{\alpha}+B\, x_{31}^{\alpha},
\end{equation}
where we note $x_{ij}^{\alpha}=(x_i-x_j)^{\alpha}$. Of all the difference vectors we could have used we omitted $x_{32}^{\alpha}$ since it is not independent ($x_{32}^{\alpha}=x_{31}^{\alpha}-x_{21}^{\alpha}$). $A$ and $B$ have to be scalar functions of the invariants $x_{21}^2$, $x_{31}^2$ and $x_{32}^2$.

Projecting both sides of the ansatz (\ref{PVansatz}) with the vectors $x_{21}^{\alpha}$ and  $x_{31}^{\alpha}$ and completing squares in the numerator of the integrand we arrive to the system of equations
\begin{align}\label{PVsystem}
& 2 A\, x_{21}^2+B(x_{21}^2+x_{31}^2-x_{32}^2)=\mathcal{B}_{\epsilon}(x_{23}^2)
-\mathcal{B}_{\epsilon}(x_{13}^2)+x_{21}^2 \mathcal{T}_{\epsilon}(x_1,x_2,x_3)
\nonumber\\
&  A (x_{21}^2+x_{31}^2-x_{32}^2)+2 B\, x_{31}^2=\mathcal{B}_{\epsilon}(x_{23}^2)
-\mathcal{B}_{\epsilon}(x_{21}^2)+x_{31}^2 \mathcal{T}_{\epsilon}(x_1,x_2,x_3)
\end{align}
where $\mathcal{B}_{\epsilon}(x_{ij}^2)$ is the regularized bubble integral defined in (\ref{Bubble}) and we know how to solve it. On the other hand $\mathcal{T}_{\epsilon}(x_1,x_2,x_3)$ is the scalar $D$-dimensional regularized star integral
\begin{equation}
\mathcal{T}_{\epsilon}(x_1,x_2,x_3)=
(l^2 e^{\gamma}\pi)^{\epsilon}\int d^D x_0\frac{1}{|x_0-x_1|^2|x_0-x_2|^2|x_0-x_3|^2}.
\end{equation}

Since the system (\ref{PVsystem}) is linear, $A$ and $B$ will be written as a complicated linear combination of Bubble integrals (which we know its solution) and scalar star integrals (which we should solve). Consider its Mellin-Barnes representation
\begin{align}
\mathcal{T}_{\epsilon}(x_1,x_2,x_3)=\frac{4\pi\ \epsilon\ (1-2\epsilon )e^{\gamma\epsilon}l^{2\epsilon}}{\Gamma(1-2\epsilon) (x_{32}^2)^{2+\epsilon}}
\int\frac{du dv}{(2\pi i)^2} &
\Gamma(-u)\Gamma(-1-\epsilon-u)\Gamma(-v)\Gamma(-1-\epsilon-v)\cdot \nonumber\\
& \cdot \Gamma(1+u+v)\Gamma(2+\epsilon+u+v)
\left(\frac{x_{21}^2}{x_{32}^2}\right)^u \left(\frac{x_{31}^2}{x_{32}^2}\right)^v,
\end{align}
where the contours go from $-i\infty$ to $i\infty$ leaving the semi-inifinite set of poles of $\Gamma(\ldots-u)$ and $\Gamma(\ldots-v)$ to the right of the contour and the semi-infinite set of poles of $\Gamma(\ldots+u)$ and $\Gamma(\ldots+v)$ to the left of the contour. Notice that there is an overall $\epsilon$ multiplying the integral. Since we are interested in the Feynman integral up to finite terms in its $\epsilon$ expansion, the overall $\epsilon$ allows us to only keep orders up to $\mathcal{O}(\epsilon^{-1})$ inside the Mellin-Barnes. One would be tempted to expand the Gamma functions inside the Mellin-Barnes, but the problem with this is that in such expansion some left poles collide with some right poles ruining the well defined contour. The way out of this problem is to deform the contour by leaving all the potentially colliding poles to one side of the contour and compensating this deformation with integrals around those poles which can be evaluated using residues. Besides those residues, the remaining Mellin-Barnes has now a well defined holomorphic $\epsilon$ expansion, but since we are interested in $\mathcal{O}(\epsilon^{-1})$ contributions from the Mellin-Barnes, that expansion is irrelevant for our aim. Thus, picking up the poles from the set 
\begin{equation}
(u,v)=\{(-1-\epsilon,-1-\epsilon),(-1-\epsilon,-\epsilon),(-1-\epsilon,0),(-\epsilon,-1-\epsilon),(0,-1-\epsilon)\}
\end{equation}
we have
\begin{align}
& \mathcal{T}_{\epsilon}(x_1,x_2,x_3)= 
\frac{2\pi e^{\gamma\epsilon}l^{2\epsilon}}{\Gamma(-1-2\epsilon) (x_{32}^2)^{2+\epsilon}}
 \left[ \Gamma^2(1+\epsilon)\Gamma(-1-2\epsilon)\Gamma(-\epsilon)
\left(\frac{x_{21}^2}{x_{32}^2}\right)^{-1-\epsilon} \left(\frac{x_{31}^2}{x_{32}^2}\right)^{-1-\epsilon}
\right.\nonumber\\
& \left. 
- \epsilon\Gamma^2(\epsilon)\Gamma(-2\epsilon)\Gamma(1-\epsilon)
\left(\frac{x_{21}^2}{x_{32}^2}\right)^{-1-\epsilon} \left(\frac{x_{31}^2}{x_{32}^2}\right)^{-\epsilon}
- \epsilon\Gamma^2(\epsilon)\Gamma(-2\epsilon)\Gamma(1-\epsilon)
\left(\frac{x_{21}^2}{x_{32}^2}\right)^{-\epsilon} \left(\frac{x_{31}^2}{x_{32}^2}\right)^{-1-\epsilon}
 \right.\nonumber\\
& \left. 
+ \Gamma(1+\epsilon)\Gamma(-\epsilon)\Gamma(-1-\epsilon)
\left(\frac{x_{21}^2}{x_{32}^2}\right)^{-1-\epsilon} 
+ \Gamma(1+\epsilon)\Gamma(-\epsilon)\Gamma(-1-\epsilon)
 \left(\frac{x_{31}^2}{x_{32}^2}\right)^{-1-\epsilon}
 \right]+\mathcal{O}(\epsilon)
\end{align}
and expanding in $\epsilon$ we obtain the symmetric result
\begin{align}
 \mathcal{T}_{\epsilon}(x_1,x_2,x_3)= &
-\frac{x_{21}^2 +x_{31}^2 +x_{32}^2}{x_{21}^2 x_{31}^2 x_{32}^2}\ \frac{2\pi}{\epsilon}
+\frac{4\pi}{x_{21}^2 x_{31}^2 x_{32}^2}\left[
(x_{21}^2+x_{31}^2-x_{32}^2)\log\frac{|x_{32}|}{l}\right.
\nonumber\\
& \left. +
(x_{31}^2+x_{32}^2-x_{21}^2)\log\frac{|x_{21}|}{l}+
(x_{21}^2+x_{32}^2-x_{31}^2)\log\frac{|x_{31}|}{l}\right]+\mathcal{O}(\epsilon).
\end{align}

Going back to the system (\ref{PVsystem}) we solve for $A$ and $B$
\begin{equation}
A=\frac{
x_{21}\cdot x_{31}\mathcal{B}_{\epsilon}(x_{21}^2)+x_{31}\cdot x_{32}\mathcal{B}_{\epsilon}(x_{32}^2)-x_{31}^2\mathcal{B}_{\epsilon}(x_{31}^2)+x_{31}^2 x_{21}\cdot x_{23}\mathcal{T}_{\epsilon}(x_1,x_2,x_3)
}{
2x_{21}^2x_{31}^2-2(x_{21}\cdot x_{31})^2
}
\end{equation}
\begin{equation}
B=\frac{
x_{21}\cdot x_{31}\mathcal{B}_{\epsilon}(x_{31}^2)+x_{21}\cdot x_{23}\mathcal{B}_{\epsilon}(x_{32}^2)-x_{21}^2\mathcal{B}_{\epsilon}(x_{21}^2)+x_{21}^2 x_{31}\cdot x_{32}\mathcal{T}_{\epsilon}(x_1,x_2,x_3)
}{
2x_{21}^2x_{31}^2-2(x_{21}\cdot x_{31})^2
}
\end{equation}
and using the results we obtained for the bubble and the scalar star integral we arrive to an impressive simplification
\begin{equation}
A=\frac{2\pi}{x_{21}^2 x_{32}^2}\left(-\frac{1}{\epsilon}+2\log\frac{|x_{21}|}{l}-2\log\frac{|x_{31}|}{l}+2\log\frac{|x_{32}|}{l}+\mathcal{O}(\epsilon)\right)
\end{equation}
\begin{equation}
B=\frac{2\pi}{x_{31}^2 x_{32}^2}\left(-\frac{1}{\epsilon}+2\log\frac{|x_{31}|}{l}-2\log\frac{|x_{21}|}{l}+2\log\frac{|x_{32}|}{l}+\mathcal{O}(\epsilon)\right).
\end{equation}
With these results and the associations $x_{21}^{\alpha}\rightarrow (\bar z-z)$ and $x_{31}^{\alpha}\rightarrow (\tau-z)$, and observing that $x_{21}^2=|z-\bar z|^2$ and $x_{31}^2=x_{32}^2=|z-\tau|^2$ we obtain
\begin{align}
\mathfrak{T}_{\epsilon}(z,\tau)= &
\frac{2\pi}{|z-\tau|^2}\left(\frac{1}{\bar z-z}+\frac{1}{\bar z-\tau}\right)\frac{1}{\epsilon}
\nonumber\\
& + \frac{4\pi}{(\tau-\bar z)^2}\left(
\frac{1}{z-\bar z}\log\frac{|z-\bar z|}{l}+\frac{2}{\tau-z}\log\frac{|z-\tau|}{l}
\right)+\mathcal{O}(\epsilon),
\end{align}
and using it in (\ref{separationIntegrals}) we finally obtain
\begin{align}
I_{Bb}^{(2,\epsilon)}(z,\tau)= & \pi\bar{p}^2\left(
-\frac{1}{\epsilon}-2\log\frac{|z-\bar z|}{l}+4\log\frac{|z-\tau|}{l}+\mathcal{O}(\epsilon)
\right)\nonumber\\
I_{Bb}^{(3,\epsilon)}(z,\tau)= & \pi{p}^2\left(
-\frac{1}{\epsilon}-2\log\frac{|z-\bar z|}{l}+4\log\frac{|z-\tau|}{l}+\mathcal{O}(\epsilon)
\right),
\end{align}
which is the result we used in the main text. Notice that after non-trivial multiple cancellations the $\mathcal{O}(\epsilon^{-1})$ contribution became independent of the distances.

\end{document}